\documentclass[12pt]{article}
\usepackage{epsfig}

\parskip=\medskipamount 
plus.3em minus.5em \abovedisplayshortskip=.5em plus.2em minus.4em
\renewcommand{\thesection}{\arabic{section}}

\catcode`@=11

\@addtoreset{equation}{section} \@addtoreset{equation}{subsection}
\def\theequation{\ifnum\value{section}=0 \arabic{equation}\ignorespaces
\else \ifnum\value{section}=-1 A.\arabic{equation}\ignorespaces
\else \ifnum\value{subsection}=0
\thesection.\arabic{equation}\ignorespaces \else
\thesection.\arabic{subsection}.\arabic{equation}\ignorespaces
                             \fi
                        \fi
                   \fi}

{\catcode`\'=\active \def'{{}^\bgroup\prim@s}}

\catcode`@=12

\newcommand{\bq}{\begin{equation}}
\newcommand{\be}{\begin{equation}}
\newcommand{\fq}{\end{equation}}
\newcommand{\ee}{\end{equation}}
\newcommand{\bqr}{\begin{eqnarray}}
\newcommand{\beqs}{\begin{eqnarray}}
\newcommand{\fqr}{\end{eqnarray}}
\newcommand{\eeqs}{\end{eqnarray}}

\newcommand{\rf}[1]{(\ref{#1})}

\def\bop#1{\setbox0=\hbox{$#1M$}\mkern1.5mu
    \vbox{\hrule height0pt depth.04\ht0
    \hbox{\vrule width.04\ht0 height.9\ht0 \kern.9\ht0
    \vrule width.04\ht0}\hrule height.04\ht0}\mkern1.5mu}

\begin{document}
\thispagestyle{empty}

\begin{flushright}
\begin{tabular}{l}
hep-th/0502058 \\
\end{tabular}
\end{flushright}

\vskip .6in
\begin{center}

{\bf Very Compact Expressions for Amplitudes}

\vskip .6in

{\bf Gordon Chalmers}
\\[5mm]

{e-mail: gordon@quartz.shango.com}

\vskip .5in minus .2in

{\bf Abstract}

\end{center}

A number theoretic algorithm is given for writing gauge theory
amplitudes in a compact manner.  It is possible to write down all
details of the complete $L$ loop amplitude with two integers, or a
complex integer.  However, a more symmetric notation requires more
integers, five or seven, depending on the type of theory.  It
is possible that in the symmetric form (or in the non-symmetric form)
that a direct (or less direct) recursive algorithm or generating
function can be developed to compute these numbers at arbitrary loop
order.  The existence of this function is implied by the recursive
structure of loop amplitudes and their analyticity, i.e. multi-particle
poles; a function requiring a finite number of computations such as a
polynomial with derivable coefficients is desired.

\vfill\break

The computation of gauge theory amplitudes has been preoccupying
modern researchers for years, and due to the complexity, requires
many tedious calculations by conventional methods.  Alternative
and more efficient methods for computing these amplitudes are
merited.  Current well used methods include string-inspired
methods, unitarity and factorization conditions, spinor helicity
and color decomposition, and the recent Calabi-Yau/gauge theory
weak-weak duality correspondence pertaining to deformations of
MHV amplitudes.  These techniques have been useful in computing tree
 and one-loop gauge theory amplitudes.

The derivation of the quantum amplitudes in the derivative expansion
are presented in \cite{ChalmersUnPub1}-\cite{Chalmers8}.

In this letter, a notation is given so that the typically
complicated kinematic expressions for loop amplitudes may be
shortened to a few lines.  These formulae, in spinor notation,
usually require a dozen pages to write down for the known one-loop
amplitudes; a few lines is a major improvement in notation.  A few
lines of integers is convenient for many reasons, including computational
ones that are typically performed for relevant cross section
calculations.

The analytic pieces of the amplitudes are considered in this work.
A modification of the notation can be made to include general
products of polylogarithms which arise at multi-loop.  Also, these
non-analytic terms are redundant in the sense that they are
derivable via perturbative unitarity.

The amplitudes are to be constructed from a small set of integers.
The integers required to specify the amplitude consist of : (1)
two in order to specify the spinor products $\langle ij \rangle$
in the numerator and denominator via a series of $n$ equations
and $m$ unknowns, required for each term in the expansion at a given
loop order, (2) one to label the internal group quantum numbers,
(3) one to label the particle spin numbers including line factors, (4)
one to describe the coefficient.

In order to specify a gauge theory amplitude the numbers
pertaining the kinematics and quantum numbers pertaining to the
$n$ particle states must be described.  The kinematics are
associated in spinor helicity notation via,

\bqr
(\sum [\sigma(i) \tilde\sigma(j) ])^{p} \ ,
\fqr
where $p$ is a number labeling the exponent of the term.  For
example, an individual term in the amplitude may contain the
factors,

\bqr
[3 4] \qquad ([ 12] + [ 48]
+ [ 49])^{-2} \ .
\label{spinorexamples}
\fqr
The amplitude contains a series of these factors in one of the
additive terms.  For an individual term in an $n$-particle amplitude,
there are many factors, which may be described in polynomial form
via

\bqr
\sum  x_{\sigma(i)} y_{\tilde\sigma(i)} + \vert p\vert = 0 \ ,
\label{spinorsum}
\fqr
with the vectors $\sigma$ and $\tilde\sigma$ describing the
indices in the inner products.  For example, the two terms in
\rf{spinorsum} are described by,

\bqr
x_1 y_2 + 1 = 0 \ ,
\fqr
and

\bqr
x_1 y_2 + x_4 y_8 + x_4 y_9 + 2 = 0 \ ,
\label{multispinorpro}
\fqr
with another minus sign required for the $2$ in \rf{multispinorpro}.
A minus sign is required in the last term in the sum; the minus sign
is specified by ordering the first term in the series via $i<j$ in
$x_i y_j$ for min(i), with min(j) pertaining to the min(i).  The
information in the series is grouped into a larger number $N$ via
another base $q$ expansion.  The inner products are described via
the expansion,

\bqr
N = \alpha x^2 + \beta x + \gamma  \ .
\fqr
The numbers $\alpha$ and $\beta$ in a well-ordered polynomial form
are,

\bqr
\alpha = \sum_{i=1}^n \sigma(i) x^{i+1} \qquad
\beta = \sum_{i=1}^n \tilde\sigma(i) x^{i+1} \ ,
\fqr
and $\gamma$ is an a priori arbitrary integer specifying the power
The numbers $\alpha$ and $\beta$ range in base $n$ up to the
maximum value $n^{n+2}$, with the coefficients $\sigma(i)$ and
$\tilde\sigma(i)$ ranging from $1$ to $n$. To uniquely specify
the kinematic factor, a number $N$ is required which must be written
in base $n^{n+2}+1$ ranging up to

\bqr
N = 0, \ldots, (n^{n+2}+1)^3 \ .
\label{baseN}
\fqr
The condition in \rf{baseN} is satisfied for $q$ less than the
maximum number.  Otherwise, a larger base is required; If for
some reason $q$ is larger than $N_{\rm max}$, which for $n=10$ is
approximately $10^{36}$ (a large pole), then a bound is required
for the amplitude, $q_{\rm max}$, and the base $q_{\rm max}$ is
used.  (This bound depends on the number of terms in the additive
expansion of the amplitude; it may proved by a partial expanson
of the denominators and the multi-particle pole information.  The
assumption that $q_{\rm max}$ is less than that in \rf{baseN} is
used.)

This number $N$ is enough to specify one of the kinematic factors
in one term of the amplitude's additive expansion.  The
specification of a complete term requires $m$ factors, which is
found via another (superseding) decomposition via a polynomial
with numbers labeling the factors,

\bqr
\sum a_i w^i .
\fqr
Assuming the maximum factor $q_{\rm max}$, and $m$ terms, the
polynomial is specified in base $N_{\rm max}$ via a number $N_i$,
for the $i$th term, bounded by $N_{\rm max}^{m+1}$.

The full expression describing the amplitude requires a series of
these numbers $N_t$, with each number parameterizing a single
additive term.  Given the three integers, $q_{\rm max}$, the
number of terms $N_{\rm terms}$, and the maximum number $N_{\rm max}$
required to specify a term (at a given order), a number $Q$ and $P$
is used via another base reduction to specify the terms, i.e.
$Q= \sum a_i x^i$ in base $P$.  The kinematics is specified by the
pair $(Q,P)$.  The bound on the number is $N_{\rm max}^{N_{\rm
terms}}$.  The complex brackets, i.e. $\langle ij\rangle$, are
specified by two more additional numbers $\bar P$ and $\bar Q$.

The remaining numbers required to specify the amplitude are the
prefactors of the individual terms and the particles' quantum
numbers.

The helicity states in pure gauge theory are described by a
vector $(\pm 1, \ldots, \pm 1)$, which may be written in
base $2$ with expansion of $0\rightarrow 1$ and $1\rightarrow -1$.
The number required is one from $0$ to $2^n$, via $\sum^{n-1} a_i
x^i$.

The group theory quantum numbers are described by another number
with a similar decomposition.  Consider U$(m)$ with $m^2$
generators.  The generators of the particle states are
parameterized by a mode expansion,

\bqr
\sum^{m^2-1} b_i x^i \ ,
\fqr
with $b_i$ a label ranging from $1$ to $m^2$.  The number is one
from $1$ to $m^2$, which requires a number of maximum
$N_R=(m^2)^{m^2}$.  The multiple trace structure, in which there are
$L+1$ traces as in,

\bqr
{\rm Tr} \prod^{a_1} T_{\sigma(j)} \cdots
 {\rm Tr} \prod^{a_{L+1}} T_{\sigma(j)}
\fqr
ordered via the particles in a permuted set, for example as $\rho=
(1,2,4,3,9,5,7,6,8)$, is described by the integers from $0$ to $L$
(or $1$ to $L+1$) as in,

\bqr
(0,0,0,1,2,1,2,1,1) \ ,
\label{tracevector}
\fqr
where the entry is the $j$th particle, and the entry directs the
matrix to the appropriate trace.  The color decomposition is well-known,
and there are $L+1$ trace structures at loop order $L$.  The vector is
parameterized by a number in base $n$ with entries $0$ to $L$ as in
$N_T = \sum c_i x^i$.  There are a maximum of $p$ trace
structures, and the number $N_T$ ranges from $0$ to $L^n$.

The prefactors of the individual terms $g_{l,a}$ require factors
of rational numbers, powers of $\pi$ with products of euler-mascheroni
constant.  There are also divergences, such as $1/\epsilon^n$.
These numbers may also be encoded in superseded numbers.

The five numbers $P+iQ$, ${\bar P}+i{\bar Q}$, $N_S$, $N_G$, and
$N_R$ together with the prefactors label the amplitude at a given loop
order $L$ and set of quantum numbers.  The lowest number of terms at
this order is labeled at $N_L(N_S,N_G,N_R)$; this number is a lower
bound on the additive complexity.  There are $N_L$ prefactors, $G_{L,a}$.

Another form of the terms of the amplitude is simply to express the
amplitude as a polynomial directly in terms of the variables $x,y$ and
$z$.  The terms

\bqr
(\sum_{i<j} \alpha_{ij} [ij])^p
\fqr
in the product are expressed as terms in the polynomial
via

\bqr
P_j(z) = p_j z_j \sum_{a<b} \alpha_{ab} x^a y^b \ ,
\label{directpoly}
\fqr
with the $\alpha_{ij}$ representing the sign of the inner products
$[ij]$ with $i<j$, and $p_j$ the exponent of the sum.  The polynomial
form, as a function of $x,y$ and $z$ may be useful in differential
equation applications.  The polynomial is

\bqr
P(x,y;z) = P_j(z)
\fqr
which has an ordering ambiguity in the coefficients $z_i$ removed
by relations between multi-point and multi-genus amplitudes such
as factorization.  The polynomials have an interpretation in a
six-dimensional algebraic variety (and associated polyhedra); this
is possibly related to the Calabi-Yau/gauge theory twistor duality
at tree level, or to a differential equation at $n$-point and loop
$l$ generating the polynomial.

The notational complexity is much simpler than writing the full
amplitude.  A typical amplitude at one-loop with six external legs
may require several pages to write on paper, whereas, perhaps only a
few hundred digits are required.

With these numbers, the natural question is how to iterate the
numbers from one order in the loop expansion to the next.  A
number of the order of $10^{1000}$ is an approximate guess.
The functional bootstrap from a set of numbers at $L$ loop order
for a given set of quantum numbers seems possible via analyticity
requirements in the general case.  The mathematical derivation of
the iteration is quite interesting, if a finite equation or
recursive approach can be given; that is, not a set of
infinite degree polynomials that label all of the numbers at
various loop orders $L$, and requiring the solutions apriori
to determine the coefficients.

An example would be a set of polynomial equations of finite
degree in the variables $P+iQ$, ${\bar P}+i{\bar Q}$, $N_S$,
$N_G$, $N_R$ with $L$ and $N_L$, or a generating function with
a finite number of initial conditions.  Given these numbers, the
coefficients $G_{L,a}$ also can be deduced via multi-loop
factorization on intermediate poles.

The generalization to further theories is direct.  The spin
reps are generalized to the numbers $(s_1,s_2, \ldots, s_n)$,
and are grouped into a single integer.  There are no further
numbers required in the multiplet description.

Known amplitude examples may be used to investigate the patterns
in the numbers.  Also, at one-loop the multi-trace subamplitudes are
redundant and can be expressed in terms of the leading trace ones;
this property should have a number theoretic description.

The full amplitudes in self-dual gauge theory described by ${\cal
L} = {\rm Tr} G^{\alpha\beta} F_{\alpha\beta}$ with $F$ the
self-dual field strength \cite{ChalmersSiegel}, i.e. the one-loop
helicity types $(+,\ldots,+,-,-)$ \cite{BCDK},

\bqr
A_{n;1} = {g^{n-2}\over 192\pi^2}  \sum_{1\leq i_1<i_2<i_3<i_4\leq n}
 {\langle i_1 i_2\rangle [i_2 i_3]\langle i_3 i_4\rangle [i_4 i_1]
\over \langle 12\rangle \langle 23\rangle \langle n1\rangle} \ ,
\label{MHVoneloop}
\fqr
are a simple gauge theory to investigate in this context (likewise
for self-dual gravity).  The amplitudes in \rf{MHVoneloop} are maximally
compressed, and independent of the non-abelian group.  The kinematics
is independent of $N_G$ and $N_R$, as well as the helicity configuration
$N_S$.  The simple integrability could manifest in a more direct fashion
on the numbers $(P+iQ,{\bar P}+i{\bar Q})$.

There are further equivalent representations of the amplitudes
in terms of integers.  There could also be some polynomial (knot)
arithmetic associated with the numbers \cite{ChalmersUnPub4}
and their relations.

\vfill\break

\end{document}